\documentstyle[prb,aps,preprint]{revtex}
\begin{document}
%\draft
\tighten
\title{Negative Magnetoresistance in the Nearest-neighbor Hopping Conduction}
\author{X. R. Wang$^1$, and X. C. Xie$^{1,2}$}
\address{$^1$ Physics Department, The Hong Kong University of Science and 
Technology, Clear Water Bay, Hong Kong}
\address{$^2$ Department of Physics, Oklahoma State University,
Stillwater, OK 74078. }
\date{\today}
\maketitle
\begin{abstract}
We propose a size effect which leads to the negative magnetoresistance 
in granular metal-insulator materials in which the hopping between two 
nearest neighbor clusters is the main transport mechanism. We show that 
the hopping probability increases with magnetic field. This is originated 
from the level crossing in a few-electron cluster. Thus,
the overlap of electronic states of two neighboring clusters increases, 
and the negative magnetoresistance is resulted. 
\end{abstract}

The magnetoresistance in various transport mechanisms has been the subject 
of many studies. For band conduction where electrons move from one place to 
another through diffusion, it is textbook knowledge that the resistance 
will increase with the strength of a strong magnetic field because electrons
can be deflected by a Lorentz force\cite{pippard}. However, the negative 
magnetoresistance has been observed and explained in many systems in the past 
two decades\cite{plee}. The negative magnetoresistance in dirty metals 
is related to weak localization phenomena\cite{plee,ps}. The negative 
magnetoresistance in the Mott's variable-range hopping(VRH) region is also
possible\cite{mksw} though the backscattering is unimportant there because 
electron states are highly localized and probability of backscattering 
is exponentially small. In the VRH case, it was pointed out\cite{boris2} 
that the average should be on the logarithm of the conductance 
so that the interference\cite{mksw,bottger} of forward tunneling paths 
is important. Replica treatment\cite{mksw} shows that the pairing of the 
tunneling paths is important to the average. A magnetic field can introduce 
phases to the tunneling paths. Thus, the pairing is weaken, the localization 
length increases, and the negative magnetoresistance is resulted. Recently, 
there is a report that the negative magnetoresistance was observed in 
Al/Al$_2$O$_3$\cite{abp} and gold\cite{belevtsev} granular materials in
which the nearest neighbor hopping is the main transport mechanism. 
There are several interesting features in the observations. The negative 
magnetoresistance occurs only in a system with large metal grains and near the 
percolation threshold\cite{abp,belevtsev}. It is unlikely that these experimental 
results can be explained either by the mechanism for the band conduction or 
by that for the VRH conduction. In this letter, we present a theory for the 
negative magnetoresistance in the nearest neighbor hopping conduction in 
granular materials.

For electron tunneling in a homogeneous material under a nonrandom 
potential, introduction of a magnetic field leads usually to a positive 
magnetoresistance\cite{sxrw}. This is due to both the destructive quantum 
interference between various tunneling paths and that a strong magnetic 
field can shrink a wavefunction through the magnetic confinement. The 
calculation on a hydrogen-like state\cite{boris3,bottger2} shows that a 
magnetic field can dramatically modify the asymptotic behavior of a 
wavefunction. A gigantic positive magnetoresistance in a doped semiconductor 
is explained by this shrinkage. However, a granular metal-insulator material 
with the nearest neighbor hopping conduction may behave differently in a 
magnetic field\cite{raikh} when the size of metallic grains are large 
enough. A magnetic field can interact with electron orbital motion. When the 
interaction is comparable with the typical level spacing (which is inverse 
proportional to the square of the grain size), a magnetic field may induce 
level crossing. The level crossing may profoundly affect the transport 
properties of granular materials. Consider two identical metallic grains 1
and 2, separated by a distance such that electrons move from one grain to 
another mainly by hopping. Let $\psi_1$ and $\psi_{2}$ be two states of each 
grain. Suppose that $\psi_1$ is right at the Fermi level while $\psi_2$ is just 
above the Fermi level in the absence of a magnetic field. Thus, the size 
of $\psi_2$ is, in general, larger than that of $\psi_1$. 
The electron hopping probability depends on the overlap 
of $\psi_{1}$ of the two grains. Increasing the strength of the magnetic 
field might cause $\psi_2$ to be at the Fermi energy, and $\psi_1$
to be above the Fermi level. In this case, 
the electron hopping probability will 
be dominated by the overlap of $\psi_{2}$ of the two grains. Therefore, the 
tunneling probability increases in the magnetic field, and  the negative 
magnetoresistance in the nearest hopping conduction should be observed.
In order to test the validity of the picture mentioned above, 
we study a simple exact solvable model for two-dimensional granular 
metal-insulator systems. We will demonstrate that the size of the 
wavefunction at the Fermi level of a metallic cluster goes through a 
step jump in a magnetic field at zero temperature. 

For the sake of simplicity, we idealize a granular metal-insulator film 
to be many identical disks -- clusters. Each cluster in a magnetic field 
is described by the following Hamiltonian
\begin{equation}
\label{Hamiltonian}
H= -\frac{1}{2m}(\vec{p}-\frac{e}{c}\vec{A})^{2}+\frac{1}{2}m\omega_{0}^{2}
(x^{2}+y^{2})
\end{equation}
where $\omega_0$ is the simple harmonic oscillator frequency parameterizing 
the metallic grains. Electrons reside in the potential. The larger 
$\omega_0$ is, the smaller the metallic grain size will be. 
It should be pointed out that, when one compares with a real 
experiment, the order of $\omega_{0}$ can be chosen in such way that 
$<r^{2}>=N\hbar / (m\omega_{0})$ is roughly equal to grain sizes, where 
$N$ is a quantity determined by the number of electrons on one grain. 
For a grain of several micrometers, $\omega_{0}$ is order of 
$10^{12}Hz$. The Schrodinger equation corresponding to (\ref{Hamiltonian}) 
is readily solved in a uniform magnetic field, B, with symmetric gauge 
$\vec A=(-By/2, Bx/2,0)$. 
The eigenfunction and eigenenergies are 
\begin{equation}
\label{wavefunction}
\psi_{n,l}(r,\theta)=\frac{1}{\sqrt{2\pi}}e^{-il\theta}(\alpha r)^{|l|}
e^{-\frac{1}{2}\alpha^{2}r^{2}}F(-n,|l|+1,\alpha^{2}r^{2})
\end{equation}
and
\begin{equation}
\label{energy}
E_{n,l}=(2n+|l|+1)\hbar \omega -l\hbar \omega_{L},
\end{equation}
where $F(a,b,x)$ is the confluent hypergeometric function,
$\alpha=\sqrt{m\omega/\hbar}$, $\omega^{2}=\omega_{0}^{2}+
\omega_{L}^{2}$, $n=0,1,2,\ldots$, and $l=0,\pm 1,\pm 2,\ldots$ with
$\omega_{L}=Be/(2mc)$. In the absence of a magnetic field, $\omega_{L}
=0$, $E_{n,l}$ is $2n+|l|+1$ fold degenerated. The degeneracies are broken
due to the second term in (\ref{energy}) which is from the
magnetic-field-orbital interaction. This is the term
which will be responsible to the level crossing. The size of
wavefunction with quantum numbers $n$ and $l$ can be found to be
\begin{equation}
\label{size}
<r^{2}>=\frac{\hbar}{m\omega}(2n+|l|+1).
\end{equation}
For a fixed
$2n+|l|$, the sizes $<r^2>$ of these wavefunctions are the same though
their energies may be different, and $<r^2>$ will 
shrink in a magnetic field because $\omega$ increases with the field. 

With a proper strength of a magnetic field, the energy of a state with a 
large $2n+|l|$, say $N+1$, can be smaller than that with $2n+|l|=N$ because
of the second term in equation (\ref{energy}). It is easy to see that the 
first crossing occurs between state of ($n=N+1$, $l=N+1$) and state of $(n=N
,$ $l=-N)$ when $(N+1)\hbar \omega - (N+1)\hbar \omega_{L}=N\hbar \omega + 
N\hbar \omega_{L}$, i.e., $(2N+1)\hbar \omega_{L} = \hbar \omega$, or $B = 
(mc\omega_{0})/(e\sqrt{N(N+1)})$. Similarly, the energy level of at least one 
of states with $2n+|l|=N+2$ is below some of $2n+|l|=N$ states when 
$B > (2mc\omega_{0})/(e\sqrt{N(N+2)})$. In general, at least one of 
$2n+|l|=N+k$ states is below some of $2n+|l|=N$ states when $B > 
(kmc\omega_{0})/(e\sqrt{N(N+k)})$. To see how the size of the wavefunction 
depends on the magnetic field. Let us assume that all $2n+|l|=N$ states 
are occupied while higher states are empty in the absence of a magnetic 
field. Then it can be shown that the highest occupied state corresponds to 
$2n+|l|=N+1$ when  $B >mc\omega_{0}/(e\sqrt{N(N+1)})$. 
According to (\ref{size}), $<r^2>$ will jump from $(N+1)\hbar/(m\omega)$ to 
$(N+2)\hbar/(m\omega)$. Increasing the magnetic field further to 
$B >3mc\omega_{0}/(e\sqrt{N(N+3)})$,  $<r^2>$ will jump to 
$(N+3)\hbar/(m\omega)$. In general, $<r^2>$ will jump to $(N+k)\hbar/
(m\omega)$ when $B$ is approximately larger than value $(k^{2}+2k+2Nk-N)mc
\omega_{0}/(e\sqrt{(N^{2}+3N)(N^{2}+2kN+2N+k^{2}+2k)})$. 
The size $<r^{2}>$ between two jumps decreases with the magnetic field 
strength because of the magnetic confinement, but the overall trend of 
$<r^{2}>$ is increasing 
with the magnetic field. However, $<r^{2}>$ can also decrease when 
the field is very strong ($\omega_{L}>>\omega_{0}$).
The increase in $<r^{2}>$ implies the larger overlap in the wavefunction 
of two hopping states if the functional form of the wavefunction
remains the same. Thus, an overall negative magnetoresistance is expected.

In order to be more precise, let us calculate the tunneling matrix 
element $t_{12}$ between two states, $\psi_{1}$ and $\psi_{2}$, of two 
metallic grains separated by a distance $d$. 
When an electron tunnels from 
an initially occupied state, say $\psi_{1}$, to the empty state, 
$\psi_{2}$, it will contribute to the hopping probability $P$ (per unit 
time). The contribution will be proportional to $|t_{12}|^{2}\exp (-\Delta
\epsilon_{12}/(KT))$, where $\Delta \epsilon_{12}$ describes the relative
energy level with respect to the Fermi energy\cite{boris2,bottger}. The 
hopping conduction can be regarded as an electron diffusion process 
in which an electron undergoes a Brownian motion from one cluster 
to another, and the diffusion constant $D$ relates to $P$ as $D=Pd^{2}$, 
where $d$ should be regarded as the average distance between two 
neighboring clusters. According to the Einstein relation, the electron 
mobility $\mu$ is given by $\mu=eD/(KT)$ which is related to the 
conductivity in the conventional way\cite{bottger}. Therefore, we can 
concentrate on how the tunneling matrix element near the Fermi energy 
depends on the magnetic field in order to study the magnetoresistance of 
the system. Let the centers of two clusters be at $(-d/2,0)$ 
and $(d/2,0)$, respectively. Assume that an electron tunnels from 
state $\psi_{1}$ of the left cluster to state $\psi_{2}$ of the right 
cluster, in the tight-binding approximation\cite{landau}, the tunneling 
matrix element $t_{12}$ is given by 
\[
\]
\begin{equation}
\label{hopping1}
t_{12}=\frac{\hbar^{2}}{m}\int_{-\infty}^{\infty} [ (\psi^{\star}_{1}
\frac {\partial \psi_{2}}{\partial x} - \psi_{2}\frac {\partial 
\psi^{\star}_{1}}{\partial x}) -
\frac{2i}{\phi_{0}}(\vec{A}\cdot \hat{x}) \psi^{\star}_{1}\psi_{2}]|_{x=0}dy,
\end{equation}  
where $\phi_0=c\hbar /e$ is the flux quanta, and $\vec{A}$ is the vector 
potential. For small $\vec{A}$, the second term will be small in comparison
with the first term. Then equation (\ref{hopping1}) can be simplified to 
\begin{equation}
\label{hopping2}
t_{12}=\frac{\hbar^{2}}{m}\int_{-\infty}^{\infty} [\psi^{\star}_{1}
\frac {\partial \psi_{2}}{\partial x} - \psi_{2}\frac {\partial 
\psi^{\star}_{1}}{\partial x}]|_{x=0}dy. 
\end{equation}
It is easy to show that $\psi_{1}$ and $\psi_{2}$ with quantum number 
$(n,l)$ can be expressed by solution (\ref{wavefunction}), in the symmetric 
gauge $\vec A=(-By/2, Bx/2,0)$, as
\begin{equation}
\label{leftwavefunction}
\psi_{1}=\exp (i\frac {e}{c\hbar}\vec{A_{0}}\cdot \vec{r}) 
\psi_{n,l}(r_{1},\theta_{1} ),
\end{equation}
\begin{equation}
\label{rightwavefunction}
\psi_{2}=\exp (-i\frac {e}{c\hbar}\vec{A_{0}}\cdot \vec{r}) 
\psi_{n,l}(r_{2},\theta_{2} ),
\end{equation}
where $\vec{A_{0}}=Bd\hat{y}/4$, $r_{1}$ and $\theta_{1}$ are the polar 
coordinates of $(r\cos(\theta)-d/2,r\sin(\theta))$, and $r_{2}$ and 
$\theta_{2}$ are the polar coordinates of $(r\cos(\theta)+d/2,r\sin(\theta))$.
The phase factors in equations (\ref{leftwavefunction}) and 
(\ref{rightwavefunction}) are due to the magnetic field. They will give the usual 
interference on the tunneling matrix element. 

At a low temperature, tunnelings between states close to the Fermi energy 
dominate the electron transport. For the simplicity, we will consider 
magnetic field dependence of the tunneling matrix element between two 
highest occupied states of two identical clusters. Therefore, the 
wavefunction should be replaced by a new one in calculating $t_{12}$ 
whenever the level crossing occurs at the Fermi level. There are 
two possible cases, two clusters with small and large separations. 
In the limit of large separation between grains, that is 
$d>>\sqrt{<r^{2}>}$, the shrinkage of wavefunction due to the magnetic 
confinement dominates over jumps in quantum number $N$. Figure 
\ref{tvsb}$a$ is the semilog plot of magnetic field dependence of 
the hopping coefficient for $d=600l_{0}$, and $N=100$ when $B=0$. 
$t_{lr}$ is in the unit of $\hbar^{2}/(2ml_{0}^{2})$, and magnetic 
field is in the unit of $B_{0}$. Overall, $t_{lr}$ decreases
with the magnetic field as expected. However, in the region of $d$ to
be several $2\sqrt{<r^{2}>}$, the jumps in quantum number dominate, and
$t_{lr}$ behaves similarly as that of $<r^{2}>$. Figure \ref{tvsb}$b$ is
the semilog plot of the magnetic field dependence of the hopping 
coefficient for $d=60l_{0}$, and $N=100$ when $B=0$. This may explain why 
the negative magnetoresistance in the nearest-neighbor-hopping region
has been observed only near the percolation threshold point\cite{abp}.
It may be interesting to point out that the transport mechanism is mainly 
through band conduction when $d\simeq 2\sqrt{<r^{2}>}$ because the overlap 
integrals between two clusters are large, and eigenstates in each cluster 
are broaded into energy bands. A rapid oscillation of $t_{lr}$ on $B$ 
is due to the oscillations of wavefunctions in the overlap region.
Of course, the validity of our formula of tunneling matrix element 
is in question in this case. 

Before summary, we address some important issues.
The physics of magneto-transport of granular metal-insulator materials 
is rich in the nearest neighbor hopping conduction. Depending on the size 
of metallic grains and the distance between two adjacent grains,  
both positive magneto-resistance and negative magneto-resistances 
are possible. 
When the inter-distance of grains is very large, the magnetic confinement 
dominates and an overall positive magnetoresistance is expected.
On the other hand, if level crossing dominates, the 
magnetoresistance is negative which happens when the potential 
barrier of two adjacent grains is small. Therefore, the negative magnetore-
sistance of a granular metal-insulator 
material should be expected only near the percolation threshold point.
Unlike the weak localization effect for a dirty metal where the 
resistance can change slightly in a weak magnetic field, this proposed 
mechanism could greatly change the resistance if conditions are right. 
Another possible mechanism for negative magnetoresistance which
is not addressed in this paper is as follows: A magnetic field may also 
increase the Fermi energy of a grain such that electrons
need to tunnel through a lower potential barrier. 
In this work, we point out that the level crossings can come 
from the magnetic-field-orbital interaction. The level crossings occur 
when this interaction is comparable with the level spacing of the system. 
Since level spacing is proportional to the $\hbar^{2}/(2mr^{2})$, where $r$ 
is the grain size, the field require to induce a level crossing depends on 
the grain size. In this model, for grain size of nanometers, the typical 
field for level crossing is of order of $10T$. Therefore, it is better 
to use material with grain size of tens nanometers to observe such jumps. 
Although we only studied a specific 
2D model, but the physics discussed in this paper
is expected to carry over to other 2D models, as well as
3D models. In particular, we use a simple harmonic oscillator 
metallic grain model to show that a magnetic field can induce the 
level crossings such that the highest occupied state in a metallic 
grain has a larger size in a magnetic field than that without a magnetic 
field. This may lead to a series of jumps in the hopping coefficient 
as one increases the magnetic field. Such jumps might have already 
been observed in a real experiment\cite{abp,belevtsev}. Of course, our 
analysis is zero temperature. In a real experiment with a finite 
temperature, these jumps are expected to be smoothed out. However, at a 
high temperature, the level crossing may become unimportant because a 
lot of states can participate in the hopping conduction. In such case, 
magnetic confinement effects may dominates. 
We have neglected Coulomb interaction in this work. The Coulomb 
energy is order of $\epsilon_{c}\sim \frac{e^2}{\kappa r}$ for 
a grain of size $r$, where $\kappa$ is the permittivity. The Coulomb 
interaction should be important when it is comparable with thermal energy
$kT$, or level spacing. It is known that Coulomb interaction can have many 
important effects on the transport properties of quantum dots\cite{dots} and 
granular metallic systems, such as Coulomb blockade and the interesting 
I-V characteristic in granular metallic systems\cite{ping}. Therefore, 
it is important to consider both the Coulomb interaction and temperature 
effects in order to make a detail analysis of real experiment results. 
Further studies are needed on the interplay of the level crossing, 
Coulomb interaction, and the temperature effects. 

In summary, we propose a level crossing effect which leads to negative
magnetoresistance in granular materials where nearest neighbor hopping
is the main transport mechanism.

We would like to thank Porf. Ping Sheng for introducing us the interesting 
problem. Many ideas in this work come from the stimulating discussions 
with him. Discussions with Prof. Zichao Gan, Prof. Qian Niu, Dr. Jiannong
Wang, Dr. A. B. Pakhomov and Mr. Shuk Chuen Ma are also acknowledged. 
This work was supported by UGC, Hong Kong, through RGC grant
and NSF-EPSCoR program.

% now the references. delete or change fake bibitem. delete next three
%   lines and directly read in your .bbl file if you use bibtex.

\begin{figure}
\caption{The magnetic field dependence of the hopping 
coefficient between highest occupied states in two adjacent grains. 
The magnetic field is in the unit of $B_{0}=mc\omega_{0}/e$, and 
$<r^{2}>$ in the figure is in the unit of $l_{0}^{2}=\hbar/(m\omega_{0})$.
(a) The semilog plot of $t_{lr}$ vs. $B$ for the case of $N=2n+|l|=100$ at 
$B=0$ and $d=600l_{0}$. $t_{lr}$ is dominated by the wavefunction shrinkage 
due to magnetic confinement. An overall positive magnetoresistance is 
observed. (b) The semilog plot of $t_{lr}$ vs. $B$ for the case of 
$N=2n+|l|=100$ at $B=0$ and $d=60l_{0}$. The magnetic field dependence 
of the hopping coefficient is dominated by the jumps in $<r^{2}>$ in this 
case, and an overall negative magnetoresistance is resulted.}\label{tvsb}  
\end{figure}
\end{document}